\documentclass[aps,pre,showpacs,twocolumn]{revtex4}
\usepackage[dvips]{graphicx}
\usepackage{psfrag}

\begin{document}

\title{Study of transients in the propagation of nonlinear waves\\
in some reaction diffusion systems}

\author{L. Giuggioli}
\affiliation{Consortium of the Americas for Interdisciplinary Science
and
Department of Physics and Astronomy,
University of New Mexico, Albuquerque, New Mexico 87131, USA}

\author{Z. Kalay}
\affiliation{Consortium of the Americas for Interdisciplinary Science
and
Department of Physics and Astronomy,
University of New Mexico, Albuquerque, New Mexico 87131, USA}

\author{V.M. Kenkre}
\affiliation{Consortium of the Americas for Interdisciplinary Science
and
Department of Physics and Astronomy,
University of New Mexico, Albuquerque, New Mexico 87131, USA}

\date{\today}

\begin{abstract}
We study the transient dynamics of single species reaction diffusion systems whose reaction terms $f(u)$ vary nonlinearly near $u\approx 0$, specifically as $f(u)\approx u^{2}$ and $f(u)\approx u^{3}$. We consider three cases, calculate 
their traveling wave fronts and speeds \emph{analytically} and solve the equations numerically with different initial conditions to study the approach to the asymptotic front shape and speed. Observed time evolution is found to be
quite sensitive to initial conditions and to display in some cases nonmonotonic behavior. Our analysis is centered on cases with $f'(0)=0$, and uncovers findings qualitatively as well quantitatively different from the more familiar reaction diffusion equations with $f'(0)>0$. These differences 
are ascribable to the disparity in time scales between the evolution of the front interior and the front tail.
\end{abstract}

\maketitle

\section{Introduction}

Reaction diffusion models are ubiquitous in science. They are commonly
employed to represent systems whose components move diffusively and whose
interaction events, described by the reaction terms, may be represented by nonlinear expressions in the macroscopic observables such as the system density.
Common examples can be found in aggregation~\cite{parisizhang},
deposition~\cite{bensimonetal}, chemical reactions~\cite{graybook}, flame combustion~\cite{zeldovich1959}, pulse propagation in nerves~\cite{scott1975} and population dynamics~\cite{okubobook,murraybook}. Extensions of reaction
diffusion studies to convective transport~\cite{giuggioliphysicad,peixotophysreve}, non-diffusive transport~\cite{manne,abk}
and spatially non-local interactions~\cite{fuentes,clercescaffkenkre} have also been studied in the recent past. Here we restrict our attention to a single-species reaction diffusion equation in 1-D in its simplest form, i.e.,
\begin{equation}
\frac{\partial u(x,t)}{\partial t}=D\frac{\partial ^{2}u(x,t)}{\partial x^{2}%
}+af(u),  \label{generalreactdiff}
\end{equation}
where $u(x,t)$ represents the density profile of the species expressed here
as a dimensionless quantity, $D$ is the diffusion constant, $a$ the growth rate
and $f(u)$ the nonlinearity. We will further assume that $f(0)=f(1)=0$, which is a property of the nonlinearity in many systems of interest.

Equations such as (\ref{generalreactdiff}) often result in propagating wavefronts. The class of reaction terms that allows this feature is rather broad but three generic types of nonlinearity may be distinguished~\cite{benguriavariationalspeed}. One type, henceforth called the \emph{first type}, corresponds to positive $f(u)$ for $0<u<1$ with $f(u)\approx u$ for $u\approx 0$. A well-known example is provided by the Fisher-Kolmogorov-Petrovskii-Piskunov (FKPP) equation~\cite{fisher1937,KPP} whose reaction term is $f(u)=u(1-u)$. Another type, henceforth called the \emph{second type}, corresponds to negative $f(u)$ for $0<u<b$ and positive $f(u)$ for $b<u<1$ such as the Zel'dovich---Frank-Kamenetsky~\cite{ZF} equation (ZF), also referred to in the literature as the reduced Nagumo equation~\cite{murraybook}, for which
$f(u)=u(u-b)(1-u)$. This change of sign in the nonlinearity is responsible to what is referred to in population dynamics as the Allee effect \cite{alleebook}; a density threshold exists below which an initial population eventually gets extinct. Recent work on pattern formation in the presence of the Allee effect may be found in ref.~\cite{clercescaffkenkre}. Finally, what we will call the \emph{third type} of nonlinearity has $f(u)$ positive for $0<u<1$ but is
nonlinear in $u$ for small $u$. Reaction diffusion equations with these kinds of reaction terms have been used, for example, in studying thermal combustion waves~\cite{zeldovich1959,calvin}, certain autocatalytic chemical reactions~\cite{metcalf} and calcium deposition in bone formation~\cite{tracqui}.

If the reaction term is of the first type, as described above, it has been shown that there exists a minimum speed for the existence of traveling fronts~\cite{fisher1937,KPP}. Such fronts are termed \emph{pulled fronts} due to their dynamics being dictated by the growth and spreading of the front tail~\cite{vansarloosone}. The value
of the asymptotic front speed can be simply obtained by calculating the spreading of small perturbations around the unstable state $u=0$.
For these nonlinearities, initial
conditions whose fronts are sufficiently steep eventually settle into
the traveling front shape with speed $2\sqrt{Daf^{\prime }(0)}$.
On the other hand, initial conditions with shallower fronts have an infinite set of possible front solutions with corresponding speeds~\cite{vansarloostwo}. These so-called \emph{pushed fronts} derive their name from the fact that
the dynamics in the nonlinear region of $f(u)$ drives the front propagation~\cite{vansarloostwo}.
In such cases, linearization techniques applied to the reaction diffusion
equation do not allow one to find the traveling front speed. The speed
selection mechanism differs greatly between the pushed and the pulled regimes. The long-time convergence to the traveling front shape and speed is argued to be exponential in the former case~\cite{vansarloostwo} and shown to be algebraic in the latter~\cite{vansarloosone}.

Whereas a great deal of attention has been given in the reaction diffusion
literature to the determination of the traveling wave front speed \cite
{benguria1994,benguriavariationalspeed}, the speed selection problem, i.e., the prediction of the asymptotic front speed from a given initial condition~\cite{aronsonweinberger,bramson}, and the long-time rate of
convergence to that speed~\cite{vansarloosreview}, the problem of the full transient dynamics has received little attention.
Our interest is in making some contribution to the study of
this problem. We ask how an initial
condition evolves to the asymptotic traveling front shape and speed. We
do so by analyzing the transient dynamics of reaction diffusion equations
whose $f(u)$'s belong to the third type.

The paper is organized as follows: in Sec. \ref{sec2} we select three specific forms of the nonlinearity and exhibit their resultant \emph{analytically obtained} traveling wave front speed and shape. Sec. \ref{sec3} is devoted to the study of their transient dynamics. Sec. \ref{conclusions} contains concluding remarks.

\section{Three specific nonlinearities and analytic traveling wave solutions}
\label{sec2}

\begin{figure}[htbp]
\centering
\resizebox{\columnwidth}{!}{\includegraphics{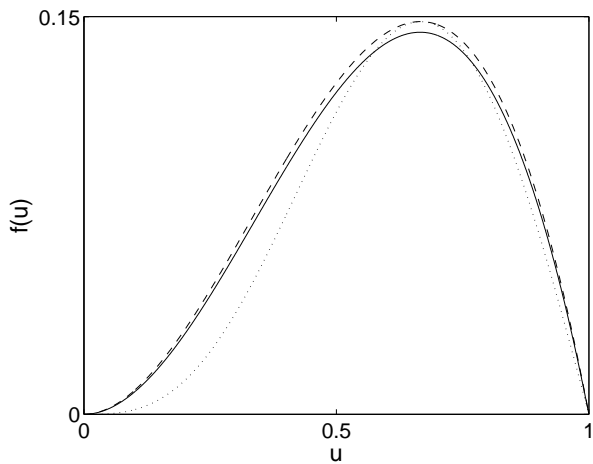}}
\caption{Plot of the nonlinearities $f(u)$ under consideration. The solid and dashed curves represent $f(u)$
quadratic in $u$ for small $u$ as given, respectively, by Eqs. (\ref{lognonlinearity}) and (\ref{ZFnonlinearity}). The dotted curve represents $f(u)$ given by Eq. (\ref{trignonlinearity}) with $\epsilon=16/(81\protect\sqrt{3})$, an
example of a nonlinearity cubic in $u$ for small $u$.}
\label{figuref}
\end{figure}

It is known~\cite{vansarloostwo,panjavansarloos} that only pushed fronts exist for reaction terms with $f^{\prime}(0)=0$. For some of these nonlinearities one can find exact analytic solutions for the traveling wave, making possible an exact description of both its shape and speed. We exploit the
existence of these pushed front solutions in studying the transient
dynamics. We consider three examples of positive nonlinearity for $0<u<1$ with $f(u)\approx u^{2}$ as well as $f(u)\approx u^{3}$ for $u\approx 0$.
These are depicted in Fig. \ref{figuref}. 

\subsection{A logarithmic nonlinearity}

Consider what we will henceforth call the \emph{logarithmic} nonlinearity,
\begin{eqnarray}
f(u)&=&(u+1)\left[ 2-\ln (2)+\ln \left( u+1\right) \right]  \nonumber \\
&&\qquad \quad\ln ^{2}\left( u+1\right) \left[ \ln (2)-\ln \left( u+1\right)
\right],
\label{lognonlinearity}
\end{eqnarray}
which is quadratic in $u$ as $u\rightarrow 0$ and decreases linearly around $u=1$. Despite its complicated appearance, we have managed to find exact
analytic solutions of the traveling wave front and speed. The procedure~\cite{vansarloosphysicad} requires the parametrization of the traveling front equation with $\partial u/\partial z$ instead of $z$. With the resulting differential equation solved exactly, an analytic value of the front speed can then be obtained. Integration of $(\partial u/\partial z)^{-1}$ and its inversion allows the determination of the exact traveling front profile. We have found the speed to be
$c=\sqrt{Da}\ln (2)$ and its traveling front shape (centered around the origin) to have the analytic form, with $z=x-ct$, 
\begin{equation}
u(z)=-1+2^{\left[1+\frac{\ln (4/3)}{\ln (3/2)}2^{z\sqrt{a/D}}\right]^{-1}}.  \label{shapelognonlinearity}
\end{equation}

\subsection{A special case of the ZF nonlinearity}

Another example of a reaction term quadratic in $u$ as $u\rightarrow 0$ may be obtained from the Zel'dovich--Frank-Kamenetsky equation by putting $b=0$, i.e., 
\begin{equation}
f(u)=u^{2}(1-u).  \label{ZFnonlinearity}
\end{equation}
The exact traveling wave front is given by 
\begin{equation}
u(z)=\frac{1}{1+e^{z\sqrt{a/2D}}},  \label{shapeZF}
\end{equation}
with $c=\sqrt{Da/2}$ as the front speed. In the following we call this $f(u)$ the \emph{quadratic} nonlinearity.

\subsection{A cubic nonlinearity}

A nonlinearity which is cubic in $u$ as $u\rightarrow 0$ can be obtained from a variety of systems including, e.g.,
\begin{equation}
f(u)=\epsilon\sin (\pi u)\left[ 1-\cos (\pi u)\right] .  \label{trignonlinearity}
\end{equation}
Given the periodicity of Eq. (\ref{trignonlinearity}), we use it in our study
only for initial conditions such that $u\leq 1$. The speed of the traveling front is given by $c=\sqrt{\epsilon\pi Da}$ and its traveling front shape is given 
\begin{equation}
u(z)=\frac{2}{\pi }\arctan \left( e^{-z\sqrt{\epsilon\pi a/D}}\right) .
\label{shapetrig}
\end{equation}
The reaction term in Eq. (\ref{trignonlinearity}) is a special case introduced to study the dynamics of the angle between the electric field and the polarization in ferroelectric chiral smectic liquid crystals \cite{vansarloos1995}. For comparison purposes
we choose $\epsilon=16/(81\protect\sqrt{3})$. This value of $\epsilon$ ensures that Eqs. (\ref{shapetrig}) and (\ref{shapeZF}) coincide at their peak $u=2/3$. In the following we will call the \emph{cubic} nonlinearity the $f(u)$ in Eq. (\ref{trignonlinearity}) with $\epsilon$ as given.

\section{Transients during the formation of the traveling wave fronts}
\label{sec3}

Our procedure for the study of transients consists of (i) assuming an
initial condition roughly in the form of a `right step' with $u=1$ as
$x\rightarrow -\infty$ and $u=0$ as $x\rightarrow +\infty$, the form of the switchover being similar in shape, but not identical, to the eventual traveling front, (ii) computing a quantity that we call the \emph{excess speed}, and (iii) studying the effects of various features of the initial conditions on the transient dynamics for the three types of nonlinearity considered. The excess speed is the amount by which the (eventual) traveling front speed $c$ is exceeded by $\upsilon(t)$, the instantaneous time rate of change of the area under the front, i.e., $d\left[\int_{-\infty}^{+\infty}dx\,u(x,t)\right]/dt$, as explained, e.g., in ref.~\cite{giuggioliphysicad}. It is a convenient quantity to focus on because it allows one to calculate how, overall, a front profile advances independently of how, on a local scale, the profile moves forward.

Our numerical integration of the reaction diffusion equation is performed via an
Adams-Bashforth-Moulton predictor-corrector method with a spatial mesh of step size 0.01 in units of $\sqrt{D/a}$, which is the characteristic length of Eq. (\ref{generalreactdiff}).
We consider the front profile as separated into a \emph{shoulder} for most of which $u=1$, a \emph{tail} which describes the final vanishing of $u$ at long distances, and an in-between \emph{interior} part.
We construct our initial conditions by modifying the asymptotic traveling shape in these three parts separately or in combination. We report results mainly for tail-modified and interior-modified initial conditions since they are the ones that show the largest effects.

In Sec. \ref{subsec3a} below, we first study initial conditions for the front obtained simply by replacing the characteristic length parameter $\sqrt{D/a}$
in the exact traveling front shape by a different value, the front shape being 
thus steeper or shallower than, but qualitatively identical to, the exact traveling front. In Sec. \ref{subsec3b} and \ref{subsec3c}, by contrast, we consider initial shapes that are different, even qualitatively, from the eventual traveling front shape. We obtain them by modifying, in the latter, only the interior portion in Sec. \ref{subsec3b}, and both the interior portion and the tail in Sec. \ref{subsec3c}. Finally, in Sec. \ref{subsec3d}, we discuss what types of initial fronts will eventually settle into the exact traveling front profiles.

\subsection{Varying the characteristic length}
\label{subsec3a}

The characteristic length $\sqrt{D/a}$ of the reaction diffusion system (\ref{generalreactdiff}) appears naturally in the exact expressions (\ref{shapelognonlinearity}), (\ref{shapeZF}) and (\ref{shapetrig}) for the traveling wave profile for the three nonlinearities, and controls the steepness of the front. In this subsection we investigate the transient occurring for initial conditions obtained by replacing $\sqrt{a/D}$ by $\xi\sqrt{a/D}$ in the respective Eqs. (\ref{shapelognonlinearity}), (\ref{shapeZF}) and (\ref{shapetrig}). The quantity we vary is $\xi$, the dimensionless ratio of the length characteristic of the initial condition to $\sqrt{D/a}$.
For values of $\xi$ larger (smaller) than 1, the initial spatial profile is shallower (steeper) than the exact traveling front.
\begin{figure}[htbp]
\centering
\resizebox{\columnwidth}{!}{\includegraphics{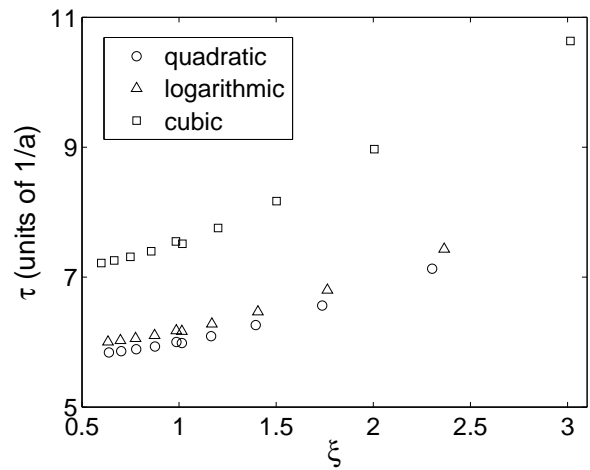}}
\caption{Long-time decay constant $\tau$ of the initial front speed plotted in units of $1/a$ for the three nonlinearities depicted in Fig. \ref{figuref}. The horizontal axis is the dimensionless characteristic length $\xi$ (see text).
The triangles, the circles and the squares correspond, respectively, to the logarithmic, quadratic and cubic nonlinearity. If $\xi>1$ ($<1$), the initial front profile is shallower (steeper) than the asymptotic profile, and the front speed decreases (increases) monotonically to the asymptotic speed.}
\label{expdecay}
\end{figure}

Our finding is that an initial condition constructed in this way gives rise to a monotonic decay of the front speed to the asymptotic value along with a change of the front steepness to the asymptotic shape. The long-time behavior appears exponential and can be fitted on a logarithmic scale by a
straight line. We calculate in this way the characteristic exponent of the decay for each of the three nonlinearities. We compare them
in Fig. \ref{expdecay} by plotting the decay time $\tau$ (reciprocal of the exponent) in units of $1/a$ as a function of the relative steepness ratio $\xi$.
It is evident from Fig. \ref{expdecay} that the decay time for the cubic nonlinearity is longer than that for the logarithmic nonlinearity. In turn, the logarithmic nonlinearity has a decay time longer than that for the quadratic nonlinearity. This hierarchy in time scales is due to the relative strength of the $f(u)$'s: the larger the value of the nonlinearity, the faster the rate of change of $u$. For a large portion of the interval $0<u<1$, the cubic nonlinearity has $u$ values smaller than the other two. Similarly, the logarithmic nonlinearity has smaller values than the quadratic nonlinearity.
Moreover, the asymmetry with respect to $\xi=1$ shows that the larger the initial deviation from the asymptotic shape, the larger the decay time
\cite{footnote1}.
The fact that shallow initial conditions can have larger differences with the asymptotic profile than steep ones explains the the increase in the transients in Fig. \ref{expdecay} as increases above the value 1.

\subsection{Modification of the interior of the front: nonmonotonic behavior}
\label{subsec3b}

\begin{figure}[htbp]
\centering
\psfrag{RCA}{$v(t)$}
\resizebox{\columnwidth}{!}{\includegraphics{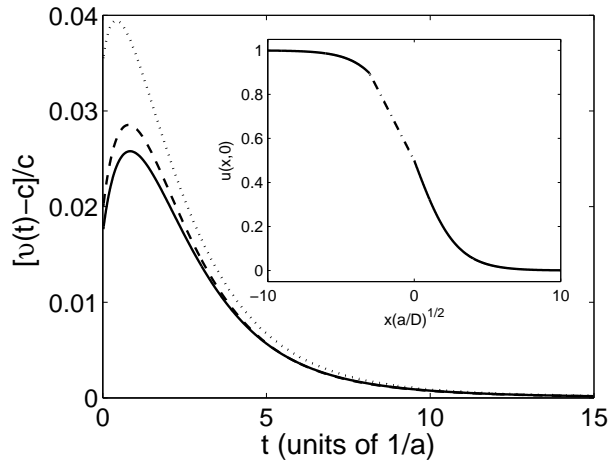}}
\caption{Nonmonotonic behavior of the relative excess speed $\left[\upsilon(t)-c\right]/c$ as function of time. The solid, dashed and dotted curve represent the evolution of the relative excess speed for the logarithmic, quadratic and cubic nonlinearities, respectively. The initial condition is constructed as follows; it is a modification of the interior of the exact traveling front profile with a straight line segment depicted in the inset as dash-dotted. Only the initial condition for the logarithmic nonlinearity is shown in the inset since the three initial conditions are very similar to each other.}
\label{perturbinit}
\end{figure}

Transients dynamics become more complex if the initial profile and the
traveling front profile are qualitatively different.
One example studied in this subsection is an initial profile obtained by modifying part of the interior of the exact traveling shape with a straight line segment. This segment starts at the coordinates $x=0$, $u=1/2$, and ends intersecting the exact traveling front profile at some point $x<0$ (see the insect of Fig. \ref{perturbinit}). In other words, the initial profile is shallower than the asymptotic one in a limited region of space.
Given that the portion of the aymptotic profile in the front interior modified to get the initial shape is rather small, one might expect a simple effect including a monotonic decay as in the previous section. Surprisingly, this does not happen. Rather than the distortion in the upper part of the interior disappearing, the shape in the lower part of the interior chnages from the exact profile and becomes shallow first.
During the transient dynamics, this situation corresponds to the appearance of a maximum in $\upsilon(t)$ as shown in Fig. \ref{perturbinit}.
At that instant, the evolution changes and $\upsilon(t)$ decays monotonically to the asymptotic value as shown in Fig. \ref{expdecay}. Notice that similar non-monotonic transients with the appearance of a maximum in $\upsilon(t)$ can be observed if the initial profile is taken as a modification of the exact profile for $x>0$, e.g., with an exponential function that is shallower than the asymptotic front shape. In other words, if the lower, rather than the upper, part of the initial front interior is made shallower than the asymptotic shape, qualitatively similar dynamics are observed.

\begin{figure}[htbp]
\centering
\resizebox{\columnwidth}{!}{\includegraphics{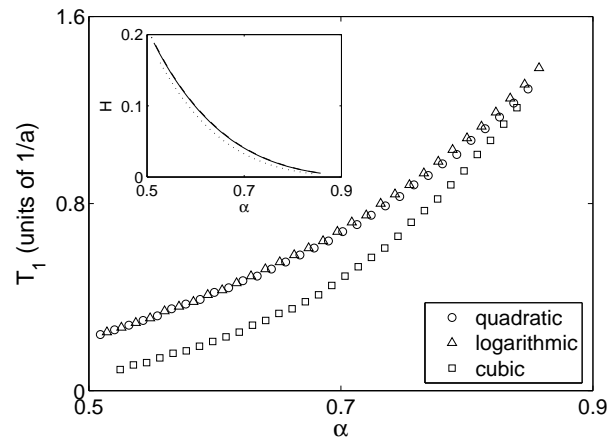}}
\caption{Plot of $T_{1}$ the transient time for the appearance of the maximum in the relative excess speed as function of the relative steepness $\alpha$ of the initial conditions displayed in the inset of Fig. \ref{perturbinit}.
The inset shows the height $H$ of the maximum of the relative excess speed as function of $\alpha$.}
\label{modifiedshoulder}
\end{figure}

We define the ratio of the slope of the straight segment to the slope of the asymptotic traveling front at $u=1/2$ as the relative steepness $\alpha$.
In Fig. \ref{modifiedshoulder}, we plot the time $T_{1}$ (in units $1/a$)
at which the maximum of $\upsilon(t)$ appears as function of $\alpha$.
The inset shows the value $H$ of the relative excess speed maximum as function of $\alpha$.
As $\alpha$ approaches 1, the distortion eventually becomes a small perturbation. The amplitude of this perturbation decays monotonically to the asymptotic profile, similarly to the long-time behavior described in Sec. \ref{subsec3a}. This explains why, for $\alpha\gtrsim 0.85$, the maximum disappears. Indeed, the inset shows that the amplitude of the maximum becomes smaller as $\alpha$ increases. The other extreme corresponds to $\alpha$ approaching 0 when the initial profile tends to flatten out for $x<0$.
In such a case, the dynamics are dominated by the reaction term since the diffusion becomes negligible. It is a simple exercise to show that a flat profile for $u>1/2$ approaches $u=1$ exponentially fast. This explains why, for $\alpha\lesssim 0.5$, the maximum in the relative excess speed also disappears.

\subsection{Modification of both the tail and the interior}
\label{subsec3c}

\begin{figure}[htbp]
\centering
\resizebox{\columnwidth}{!}{\includegraphics{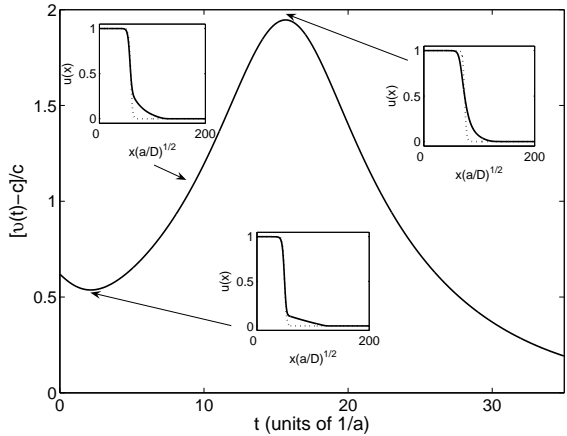}}
\caption{Relative excess speed in the case of the nonlinearity given by Eq. (\ref{lognonlinearity}) for an initial condition constructed as explained in the text. The various insets shows how the evolving front is different from the exact traveling front shape (dotted line) at different times. The exact traveling front is plotted by making the two curves cross each other at $u=1/2$.}
\label{modexactsteeptail}
\end{figure}

A common characteristic of the three nonlinearities we study is $f'(0)=0$. This implies that the dynamics in the tail of the front are slower than the dynamics in the front interior. Therefore, we may expect an even more complex transient if the tail as well as the interior of the front are modified. We show our numerical results for this situation in Fig. \ref{modexactsteeptail} for the logarithmic nonlinearity. The
initial condition is chosen as follows. The lower part of the interior of the exact traveling front is cut off at an amplitude $A=0.1$ by a straight line whose length depends on its steepness. This cut-off thus creates a tail which is infinitely steep.
At the same time the rest of the front interior is modified from the shape
of the exact solution by increasing the characteristic length $\sqrt{D/a}$ as in Sec. \ref{subsec3a}. In other words, the initial condition has an interior part which is shallower than the exact traveling front with its lower part modified with a straight line. Such an initial condition allows one to observe three different transient regimes.
\begin{figure}[htbp]
\centering
\resizebox{\columnwidth}{!}{\includegraphics{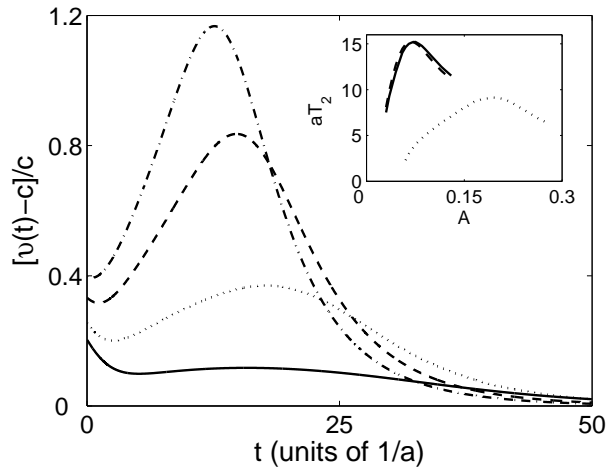}}
\caption{Plot of the relative excess speed for the logarithmic nonlinearity as function of time for four initial fronts with different value of $A$ (see text). The solid, dotted, dashed and dash-dotted curves correspond to  $A=0.04$, 0.06, 0.1 and 0.12, respectively.
The inset shows how the delay time $T_{2}$ between the minimum and the maximum changes as function of $A$ for the three nonlinearities.}
\label{transition}
\end{figure}
The first is associated with the time necessary for
the interior part of the front to reduce its steepness and become
essentially indistinguishable from the traveling front shape (see the center inset in Fig. \ref{modexactsteeptail}. The second transient is created by the fast upwards propagation of the shallow straight line profile (see the top left inset in Fig. \ref{modexactsteeptail}) present initially in the lower part of the front interior. These dynamics are similar to the one observed in Figs. \ref{perturbinit} and \ref{modifiedshoulder}:
eventually the interior front shape is converted into a shallow profile. When that happens once again $\upsilon(t)$ reaches a maximum. The final transient regime then corresponds to the monotonic decay to the asymptotic speed as shown earlier in Fig. \ref{expdecay}. Notice that, even though the initial front has a tail with compact support, i.e., is infinitely steep, that extreme steepness
is not transferred to the front interior. On the contrary, it is the shallow interior that eventually makes the interior profile shallower than the traveling wave shape during the transient.

The appearance of the two extrema is also studied by selecting initial conditions whose interior parts have a given characteristic length larger than the asymptotic traveling front, and whose right-most portion is made up of a straight line segment proceeding from $u=A$ to $u=0$. We keep the projection of the straight segment on the $x$-axis fixed and vary $A$.
In the main part of Fig. \ref{transition}, we display four curves (solid, dotted, dashed and dash-dotted) with, respectively, increasingly larger values of $A$. The effect of the time scale disparity between the tail and front dynamics can be appreciated from Fig. \ref{transition}. At first, as $A$ increases, the time lag between the minimum and maximum also increases. However, beyond a certain value of $A$, a larger portion of the shallow profile sees a rapidly increasing shape of the nonlinearity $f(u)$. The evolution of the initial front thus gets increasingly faster as $A$ increases, reducing the time necessary for the profile to become shallow overall. This explains why the maximum of the dashed and dash-dotted curves move to the left. The inset of Fig. \ref{transition} shows the variation of this time lag called $T_{2}$ as function of $A$ for the three nonlinearities. This transition is visible in all three cases but is more pronounced in the logarithmic and quadratic nonlinearities.

\subsection{Initial fronts with long tails}
\label{subsec3d}

Further studies of the transient dynamics we have carried out have shown that the types of initial fronts that eventually settle into the traveling wave shape is quite large. For the three nonlinearities considered, initial fronts whose tails decay as a power law with exponent $\gtrsim 3$, approach the asymptotic speed $c$ within the accuracy of our numerical scheme \cite{footnote2}.
For shallower initial tails, the asymptotic speed seems to settle to larger values indicating that the basin of attraction of the traveling front solution not only contains initial fronts whose tail is steeper but also those that are much shallower than the asymptotic ones. This is different from the class of nonlinearities with $f'(0)>0$ in the pushed regime. In those cases, if the initial tail decays faster than an exponential whose characteristic length depends on $f(u)$, the front speed converges to the lowest (pushed) asymptotic traveling speed. On the other hand, any initial conditions whose tail is shallower than the above exponential will converge to a larger speed value~\cite{vansarloosphysicad}.
This is not the case in the nonlinearities we have studied since the front interior dynamics is faster than the tail dynamics. The initial shape of the front interior is thus more important than the initial tail shape. Our various examples have shown that the initial front adjusts itself to the traveling front depending mostly on the shape of the interior part of the front.
The initial tail seems to determine the asymptotic speed only if it vanishes at long distances with a shallow power law profile.

\section{Concluding remarks}
\label{conclusions}

We have studied the transient dynamics of reaction diffusion systems whose reaction term $f(u)$ is characterized by being positive for $0<u<1$ with $f'(0)=0$. Specifically, we have chosen two nonlinearities whose $f(u)$ is quadratic and one whose $f(u)$ is cubic for $u\approx 0$. We have found analytic expressions for the corresponding traveling front shape and speed. With the help of these expressions, we have analyzed how certain types of initial conditions evolve in time. Our initial conditions are simple modifications of the exact traveling front shape. Monotonic relaxation to the traveling front speed is observed if the initial front is qualitatively similar to the asymptotic profile, being constructed from the latter only by varying the characteristic length scale of the traveling wave front. Nonmonotonic behavior instead may be observed when the initial conditions are qualitatively different from the asymptotic profile. In such cases, the transient dynamics is dominated by the evolution of the interior part of the front since the tail evolution is much slower due to the nonlinear nature of $f(u)$ for $u\rightarrow 0$. If part of the interior is shallower (steeper) than the asymptotic traveling front, the shallowness (steepness) is transferred to the whole interior profile, initially. Subsequently the speed then decays monotonically to zero with the long time behavior being exponential.

Many reaction diffusion systems, including the well-known FKPP case \cite{fisher1937,KPP}, have $f'(0)\not=0$, which means that,
if $u$ is the density of a species, growth can occur without the interaction
of members of the species. On the other hand, our analysis in this paper has centered on cases in which $f'(0)=0$, which represents situations in which even at low concentration, interaction (nonlinear term in $u$) is essential to growth. We find major differences relative to systems with $f'(0)>0$. In these latter systems, in the pushed regime, the initial fronts that eventually settle into the asymptotic shape are generally those whose tail decays in space exponentially with a characteristic length smaller than a value dependent on the form of $f(u)$. In our investigation, on the other hand, the initial conditions that settle into the traveling front can have tails as shallow as power laws. Given the time scale disparity between the dynamics of the tail and of the interior, the tail evolution can be said to be slaved to the interior evolution. For this reason, the tail does not affect the speed selection mechanism unless it is extremely shallow. As the integer $n$ increases in reaction terms with $f(u)\approx u^{n}$ close to $u=0$, the disparity in time scales grows bigger and we expect the basin of attraction for the initial fronts to get larger. With the reaction terms that we have used it is difficult to verify numerically that this is indeed the case. However, we can easily make a comparison with a nonlinearity whose tail dynamics are even slower than in three cases we have studied. By taking the ZF nonlinearity with $b\not=0$, we clearly see that initial conditions that decay with power laws of exponent $\gtrsim 2$ approach the traveling front profile.

In this work we have only scratched the surface of the important topic of transient dynamics. We hope that our study will motivate further research on the formation of the traveling front profile in reaction diffusion systems.

\begin{acknowledgments}
This work was supported in part by the NSF under grant no.
INT-0336343, by NSF/NIH Ecology of Infectious Diseases under grant no.
EF-0326757, by the Program in Interdisciplinary Biological and Biomedical Sciences at UNM funded by the Howard Hughes Medical Institute, and by DARPA under grant no. DARPA-N00014-03-1-0900.
\end{acknowledgments}


\begin{thebibliography}{99}
\bibitem{footnote1} We point out that the determination of $\tau$ in the close neighbourhood of $\xi=1$ presents almost insurmountable numerical difficulties. Clearly, $\tau\rightarrow 0$ as $\xi\rightarrow 1$, from the definition of the quantities involved: this is not apparent in Fig. \ref{expdecay}.
\bibitem{footnote2} Here and throughout this paper we have taken $\upsilon$ to approach $c$ if the differences is less than $10^{-7}$ in units of $\sqrt{Da}$.
\end{thebibliography}
\end{document}